\documentclass{appolb}
\usepackage{graphicx}

\def\noeq#1{(\ref{#1})}
\def\1eq#1{Eq.~(\ref{#1})}

\def\2eqs#1#2{Eqs.~(\ref{#1}) and~(\ref{#2})}
\def\3eqs#1#2#3{Eqs.~(\ref{#1}),~(\ref{#2}) and~(\ref{#3})}

\def\fig#1{Fig.~\ref{#1}}

\begin{document}

\title{Nonperturbative effects of divergent ghost loops
\thanks{Presented by D.B. at the Workshop ``Excited QCD 2014''. Bjelasnica, Sarajevo, February 2-8, 2014.}}

\author{Daniele Binosi, David Iba\~nez
\address{European Centre for Theoretical Studies in Nuclear Physics and Related Areas (ECT*) and
Fondazione Bruno Kessler,
Villa Tambosi, Strada delle Tabarelle 286, I-38123 Villazzano (TN) Italy
}}

\maketitle

\begin{abstract}
We report on a recently unveiled connection at the nonperturbative level between the masslessness of the ghost, the precise form of the gluon propagator in the deep infrared, and the divergences observed in certain kinematic limits of the three-gluon vertex.
\end{abstract}

\medskip

In recent years a detailed and systematic scrutiny of the nonperturbative properties of the fundamental Yang-Mills' Green's function has been carried out within both continuum as well as discretized methods.
As a result of these efforts, our understanding of the infrared (IR) sector of these theories has advanced considerably; in particular, there is nowadays agreement that in the Landau gauge, the gluon propagator $\Delta(q^2)$ saturates in the IR to a constant (non-vanishing) value, while the ghost propagator $D(q^2)$ diverges like the one of a free particle (and therefore, in this case, it is the ghost dressing function $F(q^2)=q^2D(q^2)$ that saturates in the IR).

In the context of the so-called PT-BFM scheme~\cite{Aguilar:2006gr,Binosi:2007pi}, these findings are explained in a rather natural way by invoking the concept of a dynamically generated gluon mass~\cite{Cornwall:1981zr}, in which case the (Euclidean) gluon propagator assumes the form  
\begin{equation} 
\Delta_{\mu\nu}(q)=P_{\mu\nu}(q)\Delta(q^2);\qquad
\Delta^{-1}(q^2) =q^2 J(q^2) + m^2(q^2),
\label{glprop}
\end{equation}
with $q^2J(q^2)$ corresponding to the  ``kinetic'' or ``wave function'' term, $m^2(q^2)$ denoting the dynamically generated mass function whose (all-order) equation has been identified in~\cite{Binosi:2012sj}, and, finally,~$P_{\mu\nu}(q)=g_{\mu\nu}-q_\mu q_\nu/q^2$ being the transverse projector.

\begin{figure}[!t]
\centerline{\includegraphics[scale=0.925]{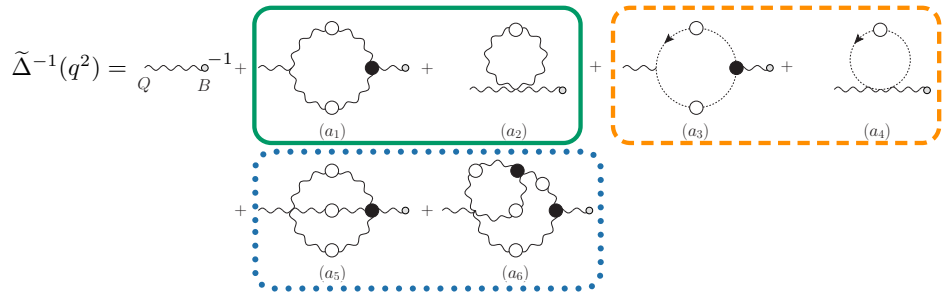}}
\caption{(color online). Diagrammatic representation of the formal relation \mbox{$[1+G(q^2)] \Delta^{-1}(q^2) = \widetilde\Delta^{-1}(q^2)$}, linking the conventional gluon propagator $\Delta$ with the propagator $\widetilde{\Delta}$ formed by a quantum gluon ($Q$) and a background one ($B$). Black (white) blobs represent fully dressed vertices (propagators); a gray circle appearing on the external legs indicates background gluons.}
\label{QB-SDE}
\end{figure}

Recently, it has been established~\cite{Aguilar:2013vaa} that the fact that the ghost remains massless at the  nonperturbative level has far reaching consequences. Specifically, the contributions to the gluon kinetic term $J(q^2)$ stemming from diagrams with ghost-loops (see Fig.~\ref{QB-SDE}) are bound to contain IR divergences. On the one hand, these divergences will modify the shape of the full propagator $\Delta(q^2)$ in the deep IR, without however interfering with its overall IR finiteness. On the other hand, higher-point functions will be also affected; in particular, certain special kinematic configurations of the three-gluon vertex, usually considered in lattice simulations, will display an IR divergent behaviour. 

The starting point for our considerations is the Schwinger-Dyson equation (SDE) for the PT-BFM gluon propagator $\widetilde{\Delta}(q^2)$ (see \fig{QB-SDE}) which is  related to  the conventional gluon propagator $\Delta(q^2)$ through the formula
\begin{equation} 
\Delta^{-1}(q^2){ P}_{\mu\nu}(q) = \frac{q^2 {P}_{\mu\nu}(q) + i\,\sum_{i=1}^{6}(a_i)_{\mu\nu}}{1+G(q^2)}.
\label{glSDE}
\end{equation}
The function $G(q^2)$ corresponds to the $g_{\mu\nu}$ form factor of a special two-point function typical of this framework~\cite{Binosi:2002ft}; in the Landau gauge, it is related to the ghost dressing function throughout the approximate equation~\cite{Grassi:2004yq}
\begin{equation}
1 + G(q^2) \approx F^{-1}(q^2),
\label{GFapp}
\end{equation}
which becomes exact in the limit $q^2\rightarrow 0$.  

Due to the special properties of the PT-BFM formalism, each of the three different boxes (continuous, dashed, and dotted line) appearing in Fig.~\ref{QB-SDE} encloses a pair of diagrams forming a gauge-invariant ({\it i.e.}, transverse) subset. As a result, one may extract from Eq.~(\ref{glSDE}) two types of \textit{individually transverse} contributions to $J(q^2)$, given by
\begin{eqnarray} 
q^2 J_g(q^2) P_{\mu\nu}(q) &=& F(q^2)\big\lbrace (a_1) + (a_2)]_{\mu\nu} + [(a_5) + (a_6)]_{\mu\nu}\big\rbrace, \label{Jg}\\
q^2 J_c(q^2)P_{\mu\nu}(q)  &=& F(q^2)[(a_3) + (a_4)]_{\mu\nu}, \label{Jc}
\end{eqnarray}
where $J_g$ ($J_c$) contains the contributions from graphs displaying gluonic (ghost) internal lines. Evidently, the complete gluon kinetic term is given by
\begin{equation} \label{Jcomplete}
J(q^2) = 1 + J_g(q^2) + J_c(q^2),
\end{equation}
with the ``1'' on the rhs coming from the tree-level term.

Now, it is natural to expect that $J_g$ will be IR-finite, since it originates from graphs with gluonic internal lines, and gluon propagators are endowed with a dynamically generated effective mass. This is to be contrasted with the corresponding contributions of graphs with ghost internal lines to $J_c$ which, due to the fact that the ghost remains nonperturbatively massless, are ``unprotected'' in the sense that there is no mass term that could tame the IR divergences they give rise to. This might be easily understood at the perturbative level: gluonic loops give rise to logarithms of the type $\log(q^2+m^2)/\mu^2$ (which are evidently finite for arbitrary momenta), while logarithms stemming from ghost loops would simply be of the type $\log q^2/\mu^2$ and therefore IR divergent.

Indeed, following the detailed analysis presented in~\cite{Aguilar:2013vaa} for the two diagrams $(a_3)$ and $(a_4)$, it can be shown that the leading contribution to Eq.~(\ref{Jc}) in the IR is given by the expression
\begin{equation} 
J_c^\ell(q^2) = \frac{g^2N}{2(d-1)}F(q^2) \int\!\frac{{\rm d}^dk}{(2\pi)^d}\,\frac{F(k)}{k^2(k+q)^2},
\label{Jdiv}
\end{equation}
with $N$ the number of colors and $d$ the space-time dimension. Since the ghost dressing function behaves as a constant in the deep IR limit $q^2\rightarrow 0$, the integral in Eq.~(\ref{Jdiv}) develops a logarithmic (linear) divergence in $d=4$ ($d=3$), as announced; of course, the product $q^2J_c^\ell(q^2)$ will be still regular as $q^2\to0$.  

\begin{figure}[!t]
\centerline{\includegraphics[scale=0.775]{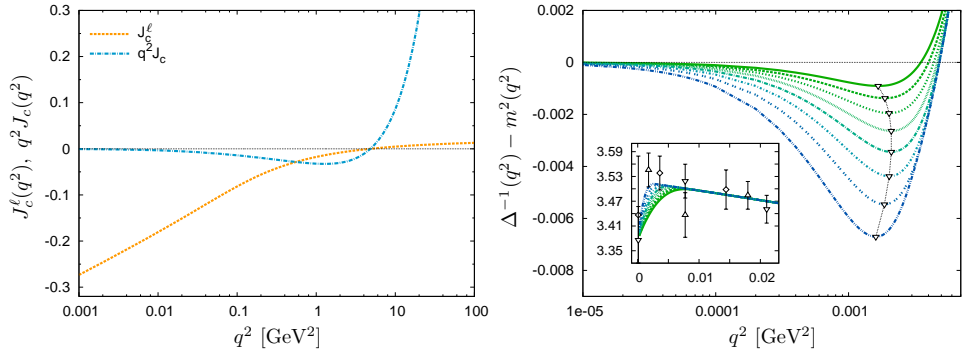}}
\caption{\label{fig:gh-Jg} \textit{(Left panel.)} The four-dimensional SU(2) ghost-loop contribution $q^2J_c(q^2)$ to the gluon kinetic term $q^2J(q^2)$ and its leading IR divergent term $J_c^\ell(q^2)$. \textit{(Right panel.)} The gluon full kinetic term $q^2J(q^2)=\Delta^{-1}(q^2)-m^2(q^2)$ displays a negative IR minimum (marked by the open down triangles) as a consequence of the IR maximum developed by the propagator (inset).}
\end{figure}

In the 4-dimensional case the aforementioned IR divergence can be clearly seen on the left panel of~\fig{fig:gh-Jg}, where we plot both the SU(2) full ghost-loop contribution $q^2J_c(q^2)$ to the gluon kinetic term, as well as its IR leading term 
$J_c^\ell(q^2)$~\noeq{Jdiv}, evaluated numerically using as input the available quenched lattice data for the ghost dressing function~\cite{Cucchieri:2010xr}.

The presence of this divergent term may be combined with known properties of the solutions of the mass equation~$m^2(q^2
)$~\cite{Binosi:2012sj}, in order to show that the derivative of the (inverse) gluon propagator
\begin{equation} 
[\Delta^{-1}(q^2)]^{\prime} = J(q^2) + q^2J'(q^2) + [m^2(q^2)]^{\prime},
\label{derprop}
\end{equation}
has a zero in the IR, which corresponds to a maximum of the gluon propagator. This is shown in the inset of the right panel of~\fig{fig:gh-Jg}, where the predicted propagator behavior is compared with the corresponding lattice data~\cite{Cucchieri:2010xr}. In the main panel we show instead that how this gluon propagator maximum translates into the presence of an IR minimum for the full gluon kinetic term $q^2J(q^2)=\Delta^{-1}(q^2)-m^2(q^2)$. 

The effect of divergent ghost loops is far from being exhausted at the two-point sector level, rather influencing the whole tower of $n$-point gluon Green's functions. For example, it turns out that it is precisely 
the gluon kinetic term that accounts for the divergent behavior of the three-gluon vertex in certain kinematic limits studied on the lattice.

Specifically, the quantity usually employed on the lattice is the ratio
\begin{equation} \label{Rproj}
R(q,r,p) = \frac{\Gamma^{(0)}_{\alpha\mu\nu}(q,r,p)P^{\alpha\rho}(q)P^{\mu\sigma}(r)P^{\nu\tau}(p)\Gamma_{\rho\sigma\tau}(q,r,p)}{\Gamma^{(0)}_{\alpha\mu\nu}(q,r,p)P^{\alpha\rho}(q)P^{\mu\sigma}(r)P^{\nu\tau}(p)\Gamma^{(0)}_{\rho\sigma\tau}(q,r,p)},
\end{equation}
in which the full vertex $\Gamma_{\rho\sigma\tau}$ is projected onto its tree-level value (indicated by a $(0)$ superscript), dividing out, at the same time, the external legs corrections. This ratio may be characterized by the modulo of two independent momenta (say $q^2$ and $r^2$) and the angle $\varphi$ formed between them. Then, in the so-called ``orthogonal configuration'', with $\varphi=\pi/2$, and the kinematical limit $r^2\rightarrow 0$, one can show~\cite{Aguilar:2013vaa} that the behaviour of $R$ in the deep IR is determined solely by $J(q^2)$
\begin{equation} \label{RIR}
R(q^2)\stackrel{q^2\to 0}{\sim}F(0)[q^2J(q^2)]'\sim F(0)
J(q^2).
\end{equation}

\begin{figure}[!t]
\centerline{\includegraphics[scale=0.765]{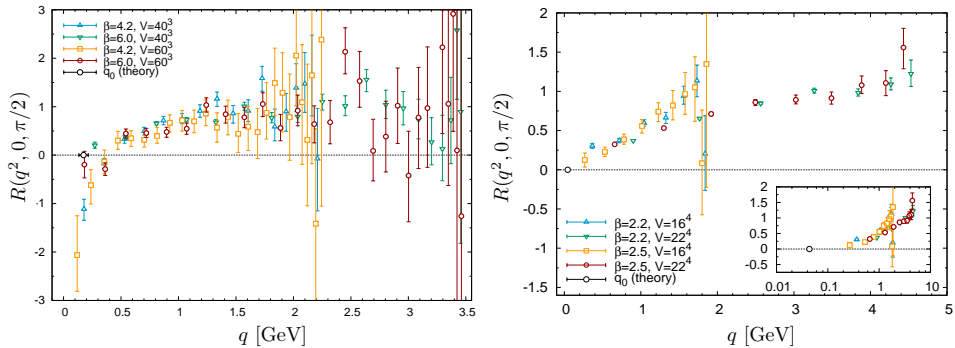}}
\caption{\label{fig:3g-R} Prediction for the zero-crossing of the SU(2) $R$-projector (black dot) compared with the corresponding lattice data in 3 (left) and 4 (right) dimensions.}
\end{figure}

According to this equation then, $R(q^2,0,\pi/2)$ should display a zero crossing at a position located around around the one in which the minimum of the gluon kinetic term appears. This is shown in~\fig{fig:3g-R} for the SU(2) gauge group in both the 3- and 4-dimensional cases, where our zero crossing estimates (at $q_0\approx380$ MeV and $q_0\approx44$ MeV, respectively) are compared with the lattice data of~\cite{Cucchieri:2008qm}. Notice that while in $d=3$ our results are in agreement with the data, in $d=4$ the zero crossing is not yet resolved by the lattice; in addition, given the very low value we get for $q_0$ in this case, simulations as the ones in~\cite{Cucchieri:2008qm} running at a lattice spacing $a\approx21$ fm and $\beta=2.5$ would require a number of lattice sites of roughly $130^4$ in order to resolve $q_0$, which does not seem currently attainable (at the moment one has $L = 22$ at most). The situation gets better for the SU(3) case, in which our method yields $q_0\approx132$ MeV; assuming that lattice simulations could be performed at $a\approx0.17$ fm and $\beta = 5.7$ (that is, in the same conditions used for simulating the gluon and ghost two-point sectors), one obtains instead a number of lattice sites of around $60^4$, which looks feasible. These results are in agreement with~\cite{Blum:2014gna}, while somewhat higher values for $q_0$ have been recently obtained in~\cite{Eichmann:2014xya} through a solution of a truncated version of the three-gluon vertex SDE.

Summarizing, we have shown that the fact that the ghost field remains nonperturbatively massless, as opposed to the gluon which acquires a dynamically generated effective mass, implies unavoidably the existence of a (negative) IR divergence in the kinetic term $J(q^2)$ of the gluon propagator. While this divergence, which originates exclusively from  diagrams involving ghost loops,  does not affect the finiteness of the gluon two-point function, its presence manifests itself in the entire tower of $n$-point gluon Green's functions. In particular, we have found that in the two-point sector the $d=4$ gluon propagator displays an IR maximum (as it happens in $d=3$), while in the three-point sector the three-gluon vertex develops an IR (negative) divergence.



\end{document}